# Accurate and robust unitary transformation of a high-dimensional quantum system

B. E. Anderson[1,3], H. Sosa-Martinez[1], C. A. Riofrío[2], I. H. Deutsch[2], and P. S. Jessen[1]

**Quantum control in large dimensional Hilbert spaces is essential for realizing the power of quantum information processing. For closed quantum systems the relevant input-output maps are unitary transformations, and the fundamental challenge becomes how to implement these with high fidelity in the presence of experimental imperfections and decoherence. For two-level systems (qubits) most aspects of unitary control are well understood, but for systems with Hilbert space dimension *d*>2 (qudits), many questions remain regarding the optimal design of control Hamiltonians[1] and the feasibility of robust implementation[2,3]. Here we show that arbitrary, randomly chosen unitary transformations can be efficiently designed and implemented in a large dimensional Hilbert space (*d*=16) associated with the electronic ground state of atomic $^{133}$Cs,[4] achieving fidelities above 0.98 as measured by randomized benchmarking[5]. Generalizing the concepts of inhomogeneous control[6] and dynamical decoupling[7] to *d*>2 systems, we further demonstrate that these qudit unitary maps can be made robust to both static and dynamic perturbations. Potential applications include improved fault-tolerance in universal quantum computation[8], nonclassical state preparation for high-precision metrology[9], implementation of quantum simulations[10], and the study of fundamental physics related to open quantum systems and quantum chaos[11].**

The goal of quantum control is to perform a desired transformation through dynamical evolution driven by a control Hamiltonian $H_C(t)$. For example, one common objective is to evolve the system from a known initial state to a desired final state. If the control task is simple or special symmetries are present, it is sometimes possible to find a high-performing control Hamiltonian through intuition, or to construct one using group theoretic methods[12]. In this letter we explore the use of "optimal control"[1] to design control Hamiltonians for tasks of varying complexity, from state-to-state maps to unitary maps on the entire accessible Hilbert space. The basic procedure is well established: the Hamiltonian $H_C(t)$ is parameterized by a set of control variables, and a numerical search is performed to find values that optimize the fidelity with which the control objective is achieved. The application of optimal control to quantum systems originated in NMR[13] and physical chemistry[1], and has since expanded to include, e. g., ultrafast physics[14], cold atoms[15,16], biological molecules[17], spins in condensed matter[18], and superconducting circuits[19].

We study the efficacy of numerical design and the performance of the resulting control Hamiltonians using a well developed testbed consisting of the electron and nuclear spins of individual $^{133}$Cs atoms driven by radiofrequency (rf) and microwave ($\mu$w) magnetic fields (Fig. 1)[16]. Our experiments show that the optimal control strategy is adaptable to a wide range of control tasks, and that it can generate control Hamiltonians with excellent performance even in the presence of experimental imperfections and external perturbations. Averaging over large samples of randomly chosen transformations, we reliably achieve fidelities that range from 0.982(2) for 16-dimensional unitary maps to 0.995(1) for state-to-state maps (errors are one standard deviation). These results represent a significant advance over current state-of-the art for systems with similar-sized Hilbert spaces[19,20]. Furthermore, given that the optimal control paradigm applies to any physical platform regardless of specifics, our work provides a template for similar advances elsewhere.

Introductions to optimal control can be found in the literature[1], and we review the method only as it applies here[4]. Typically, one starts with a control Hamiltonian of the form $H_C(t) = H_0 + \sum_j b_j(t) H_j$, chosen so it can generate all possible unitary maps and renders the system "controllable". The *control waveforms* are coarse grained in time, $\{b_j(t)\} \to \{b_j(t_k)\}$, to yield a discrete set of control variables. Given a target unitary map $W$ acting in the system space $\mathcal{H}$, one can search for a set $\{b_j(t_k)\}$ that minimizes the Hilbert-Schmidt distance $\|W - U(T)\|$, where $U(T)$ is the propagator driven by $H_C(t)$ during the time $T$. If the overall phase of $W$ is unimportant, one can instead maximize the "standard" fidelity $\mathcal{F}_S = |Tr[W^\dagger U(T)]|^2/d^2$. Similarly, a map $W_{if}$ from an initial subspace $\mathcal{H}_i$ to a final subspace $\mathcal{H}_f$ can be obtained by optimizing the fidelity $\mathcal{F}_S = |Tr[W_{if}^\dagger P_f U(T) P_i]|^2/p^2$, where $p$ is the dimension of the subspaces, and $P_i$ and $P_f$ are the projectors onto them. In practice $H_C(t)$ may depend on additional parameters $\Lambda = \{\lambda_i\}$ that are imperfectly known. In that case one can search for *robust* control waveforms by maximizing the average fidelity $\bar{\mathcal{F}}_S = \int_\Lambda \mathcal{P}(\Lambda)\mathcal{F}_S(\Lambda)d\Lambda$, where $\mathcal{P}(\Lambda)$ is the

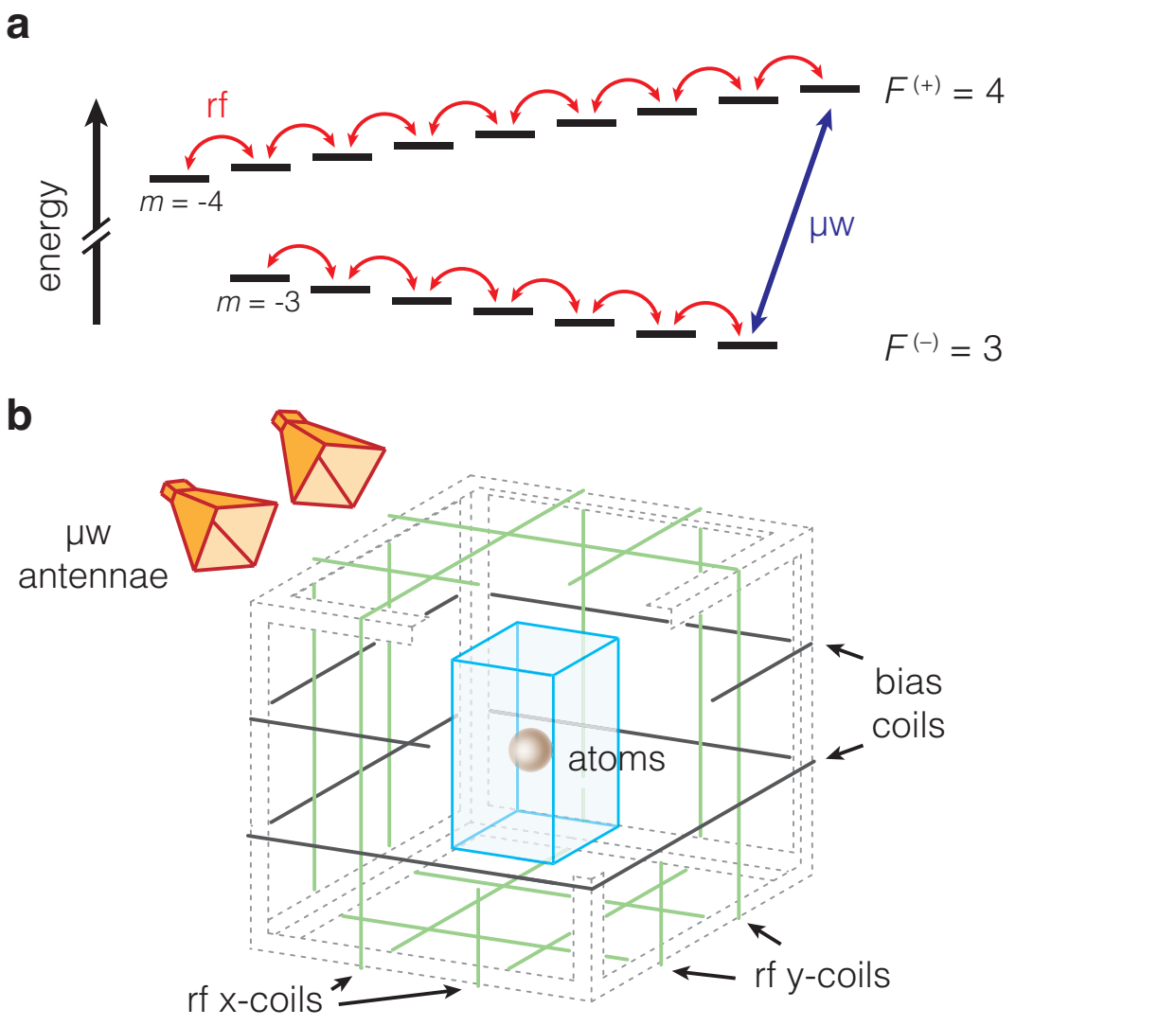

**Figure 1 | Hyperfine manifold in the electronic ground state of $^{133}$Cs. a,** Nuclear and valence electron spins combine to form a total spin with possible quantum numbers $F^{(\pm)} = 3,4$. The hyperfine interaction splits the $F^{(\pm)}$ manifolds, and the remaining Zeeman degeneracies are lifted with a bias magnetic field. The system is controlled with rf (red) and $\mu$w magnetic fields (blue). **b,** Schematic of the experimental setup. Atoms are prepared in a vapor-cell Magneto-Optic Trap, surrounded by bias and rf coils and $\mu$w antennae.

Center for Quantum Information and Control. [1]College of Optical Sciences, University of Arizona, Tucson, AZ 85721. [2]Department of Physics and Astronomy, University of New Mexico, Albuquerque, NM 87131. [3]National Institute of Science and Technology, Gaithersburg, MD 20899.

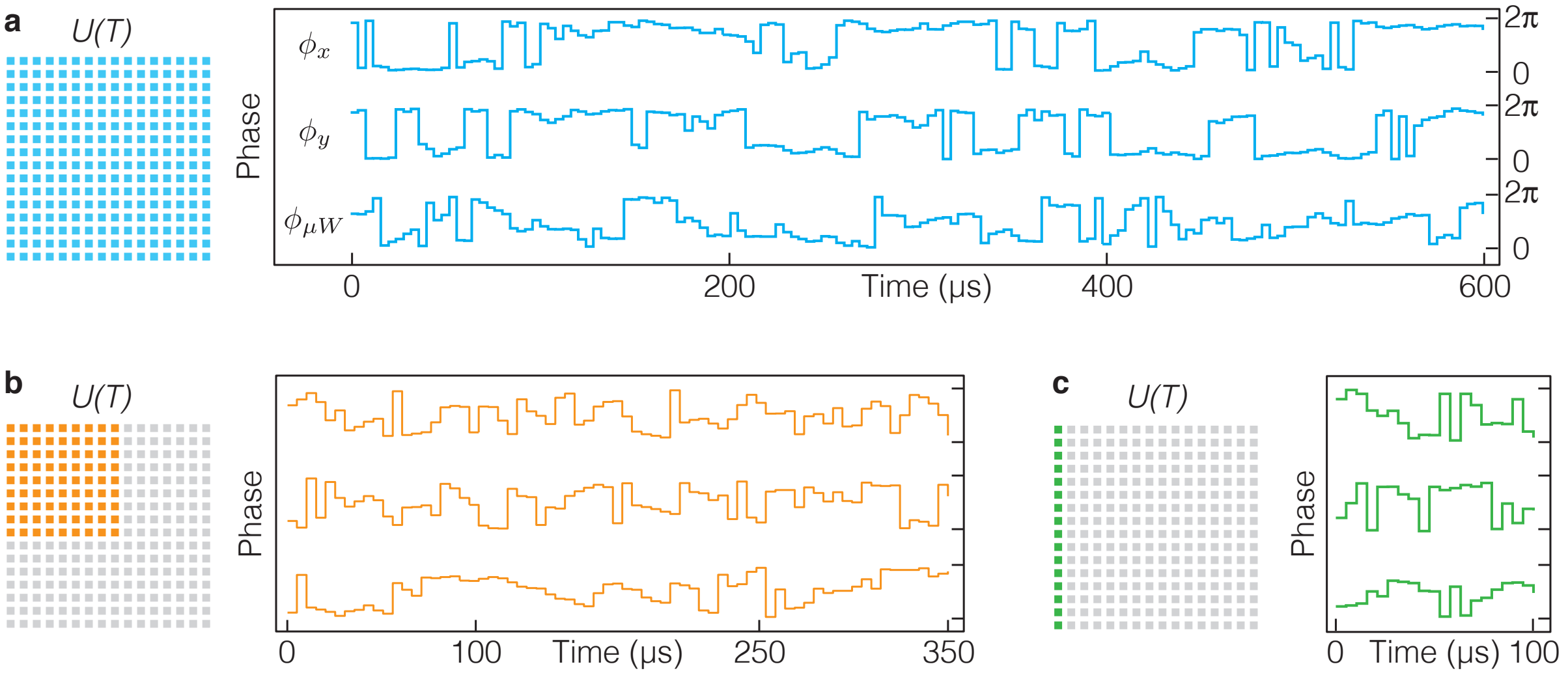

**Figure 2 | Phase modulation waveforms for control tasks of varying complexity.** **(a)** For a unitary map on the entire Hilbert space $\mathcal{H}$ every element of $U(T)$ is constrained (blue), and the control waveforms must have at least $d^2-1=255$ independent phases. In our setup a control time $T=600\mu s$ and phase step duration $\delta t=4\mu s$ is close to optimal (see Fig. 4), corresponding to a total of 450 phases. **(b)** A unitary map on the *p*-dimensional subspace consisting of the $F^{(+)}$ manifold ( $p=9$ ) constrains a $p\times p$ block of $U(T)$ (orange). The control waveforms must therefore contain at least $p^2-1=80$ phases, and we have successfully used a total of 210. **(c)** A State-to-state map in $\mathcal{H}$ constrains a single column of $U(T)$ (green). The control waveforms must contain at least $2d-2=30$ phases, and we have successfully used a total of 60.

probability that the parameters take on the value $\Lambda$ , and $\mathcal{F}_S(\Lambda)$ is the corresponding fidelity. If the parameters vary with time, one can average over an ensemble of histories, $\Lambda=\{\lambda_i(t)\}$ , and search for control waveforms with built-in dynamical decoupling[7]. Robustness is essential for quantum control in real-world scenarios, but until now little has been known about its feasibility in large Hilbert spaces.

The structure of the $^{133}$Cs electronic ground state reflects the addition of electron and nuclear spins to form the total hyperfine spin, $\mathbf{F}=\mathbf{S}+\mathbf{I}$ . The Hilbert space consists of two manifolds with spin quantum numbers $F^{(\pm)}=I\pm S=7/2\pm 1/2=3,4$ , and has overall dimension $d=(2S+1)(2I+1)=16$ (Fig. 1a). This system is controllable with a static bias field along *z*, a pair of phase modulated rf magnetic fields along *x* and *y*, and a phase modulated $\mu$w magnetic field coupling the states $|F^{(\pm)},m=F^{(\pm)}\rangle$ [4]. Following the general approach outlined above, we use control waveforms $\{\phi_{rf}^{(x)}(t_k),\phi_{rf}^{(y)}(t_k),\phi_{\mu w}(t_k)\}$ that correspond to piece-wise constant phase modulation. In our setup the dominant source of uncertainty in $H_C(t)$ is the magnitude of the static bias field. We have found empirically that robust control can be achieved by maximizing a two-point average, $\bar{\mathcal{F}}_S=\frac{1}{2}[\mathcal{F}_S(B_0+\delta B)+\mathcal{F}_S(B_0-\delta B)]$ , where $B_0$ is the nominal bias field and $\delta B$ is a static offset that characterizes its spatial inhomogeneity. Given some target unitary map *W*, we start by specifying the overall control time *T* and phase-step duration $\delta t$ and then generate a random initial guess for the control phases. This guess seeds a gradient ascent algorithm, which eventually converges on a set of control waveforms that correspond to a local maximum of the fidelity. For appropriate *T* and $\delta t$ (see below) we find that a small number of initial guesses (~10) almost always lead to at least one set of control waveforms with theoretical fidelity ≥0.999. This is consistent with the benign character of quantum control landscapes found in theoretical studies[21,22].

As expected, different quantum maps require control waveforms of different complexity. Figure 2a shows robust control waveforms designed for a randomly chosen unitary map on the entire 16 dimensional Hilbert space $\mathcal{H}$ . In this case every element of the matrices $U(T)$ and $W$ must be identical. A *d*-dimensional unitary matrix *W* in the group SU(*d*) requires $d^2-1=255$ real numbers to specify, and thus the control waveforms must contain at least that many independent phases. In practice a substantially larger number is needed to achieve robust control. Similarly, Figs. 2b&c show control waveforms for a unitary map on the 9-dimensional subspace of the $F^{(+)}$ manifold, and for a state-to-state map. These examples illustrate how control waveforms can be simpler and shorter as the constraints on $U(T)$ are relaxed.

Our laboratory setup (Fig. 1b) has been described in detail elsewhere[16]. The basic experimental sequence is performed in parallel on an ensemble of a few million Cs atoms, and consists of initial state preparation, implementation of a quantum map, and finally a measurement of the output populations in the hyperfine magnetic sublevels $|F,m\rangle$ . In principle one can reconstruct a quantum map through process tomography, but in practice this procedure is too complex and error prone for our needs here. We rely instead on randomized benchmarking, a protocol developed for qubits and quantum gates[5], and recently generalized to state-to-state maps in our system[16]. An example of benchmarking data for a random sample of 16-dimensional unitary maps is shown in Fig. 3, from which we estimate "benchmark" fidelities $\mathcal{F}_B=0.982(2)$ and $\mathcal{F}_B = 0.970(1)$ for robust and non-robust control waveforms, respectively. Similar data yield a fidelity $\mathcal{F}_B = 0.984(2)$ for unitary maps on the $F^{(+)}$ subspace, a fidelity $\mathcal{F}_B = 0.995(1)$ for maps between randomly chosen 2-dimensional subspaces $\mathcal{H}_i$ and $\mathcal{H}_f$, and a fidelity $\mathcal{F}_B = 0.995(1)$ for state-to-state maps, in all cases using robust control waveforms. Note that the measured $\mathcal{F}_B$ lie consistently below our design goal of ≥0.999, the more so for complex tasks that require longer control waveforms. This is consistent with errors from imperfect control fields and external perturbations that accumulate over time.

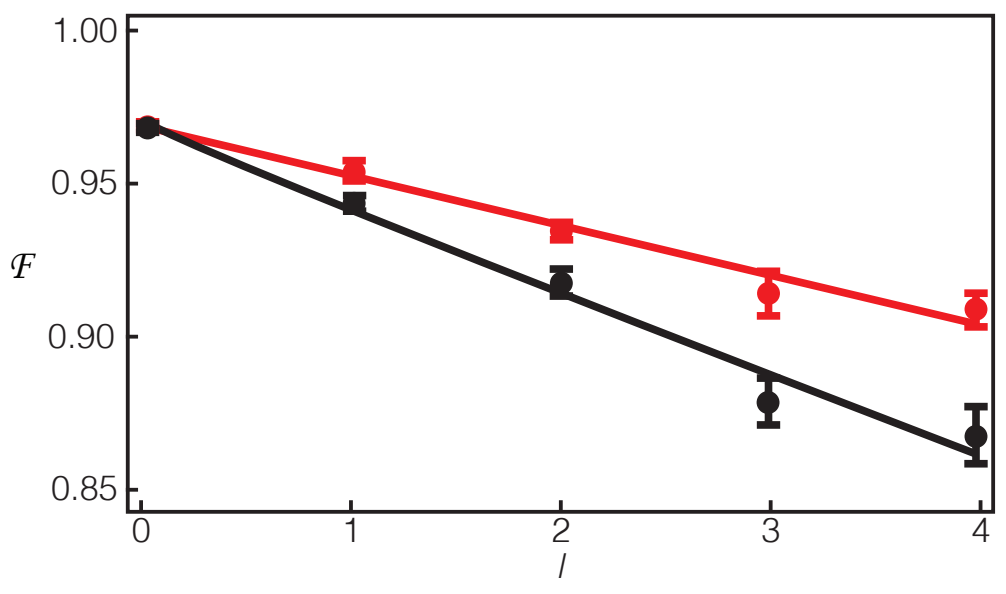

**Figure 3 | Randomized benchmarking.** Quantum maps from a representative set are combined into random sequences, and the overall input-output fidelities $\mathcal{F}(l)$ for random input states are measured as a function of sequence length *l*. The data shown here is for 16-dimensional unitary maps on the entire space $\mathcal{H}$ , implemented with robust (red) or non-robust (black) control waveforms. Each data point represents an average of 10 different sequences; error bars are ± one standard deviation of the average. Lines are fits from which the benchmark fidelity $\mathcal{F}_B$ is determined. For details see Methods.

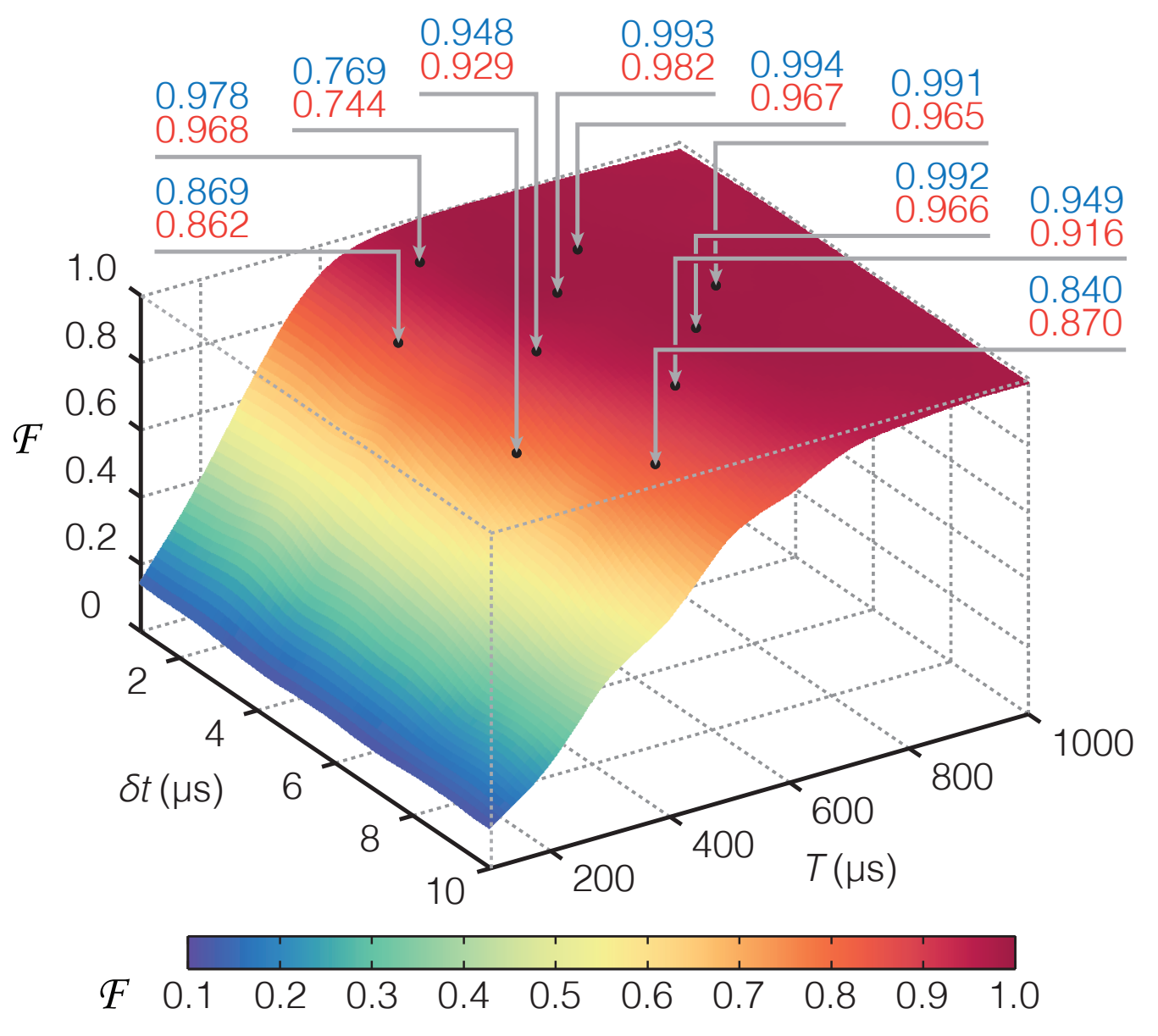

**Figure 4 | Fidelity as function of control time and phase step duration.** Average fidelity $\mathcal{F}_S$ reached by a random set of 16-dimensional unitary maps, as function of the control time $T$ and phase step duration $\delta t$. Numbers indicate average fidelities $\mathcal{F}_S$ (blue) and benchmarked fidelities $\mathcal{F}_B$ for a few discrete combinations $(T, \delta t)$.

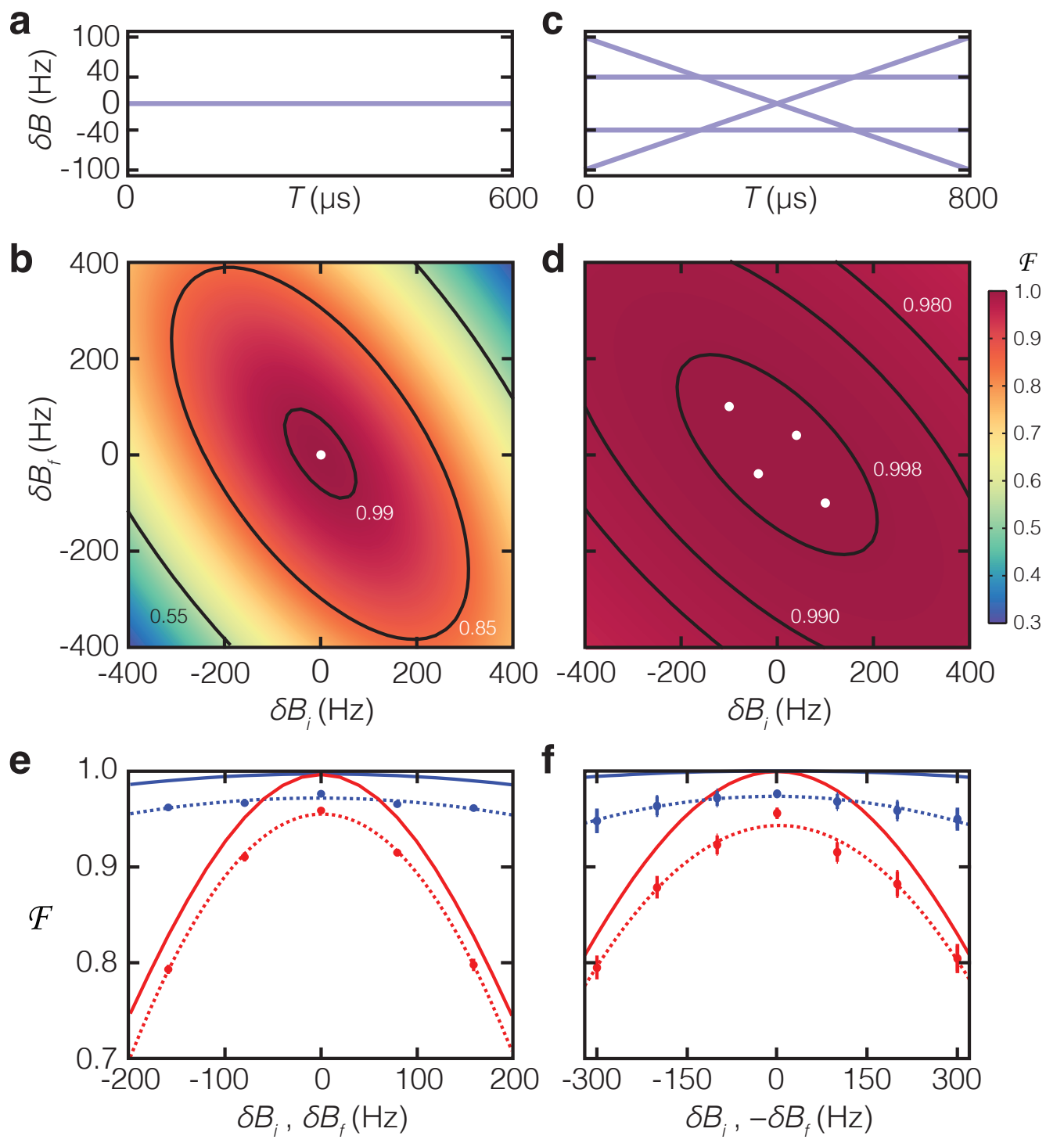

**Figure 5 | Fidelity of robust vs. non-robust control waveforms. a**, Bias field variation $\delta B(t)$ assumed in the design of non-robust control waveforms for a set of unitary maps. **b**, Average fidelity $\mathcal{F}_S$ predicted for these non-robust control waveforms when the actual $\delta B(t)$ changes linearly from $\delta B_i$ to $\delta B_f$. The white dot corresponds to the variation in **a**. **c**, Bias field variations $\delta B(t)$ used for the four-point average that goes into the design of robust control waveforms for the same set of unitary maps. **d**, Average fidelity $\mathcal{F}_S$ predicted for these robust control waveforms as function of the actual $\delta B_i, \delta B_f$. White dots correspond to the variations in **c**. **e**, Measured and predicted fidelities for robust (blue) and non-robust (red) control waveforms, along the diagonal $\delta B_i = \delta B_f$. **e**, Same along the diagonal $\delta B_i = -\delta B_f$. Data points in **e** and **f** show the average $\mathcal{F}_B$ for the set of maps; error bars are ± one standard deviation of the average. Dashed lines are parabolic fits to guide the eye. Solid lines show the predicted $\mathcal{F}_S$. Magnetic fields are given in units of Larmor frequency.

The unitary maps of Fig. 3 were implemented with control waveforms of phase step duration $\delta t = 4\,\mu s$ and overall control time $T = 600\,\mu s$. These values were identified as near-optimal after a systematic computer-numerical and laboratory exploration. Figure 4 shows the average fidelity calculated for that same set of maps when using control waveforms with a range of $(T, \delta t)$. Also shown are the benchmark fidelities measured for the maps at a few discrete points. The most notable feature is the existence of a high fidelity plateau, dropping off sharply when $T$ is too short to accommodate the required dynamical evolution, or when the ratio $T/\delta t$ does not allow for a sufficient number of control phases. In our experiment the relevant timescales for both $T_{\min}$ and $\delta t_{\min}$ are set by the rf Larmor frequencies (25 kHz) and the $\mu$w Rabi frequency (27.5 kHz), with the latter limited also by the modulation bandwidth of the rf and $\mu$w fields. Less dramatically, the experimental data shows a small decline in fidelity for $T >> T_{\min}$, due to accumulating errors from imperfections and perturbations. Based on Fig. 4, the optimum combination in our experiment appears to be around $\delta t = 4\,\mu s$ and $T = 600\,\mu s$ as stated above. Similar analyses show decreasing $T_{\min}$ for simpler tasks; we find optimal control times of 350$\mu$s for unitary maps on the $F^{(+)}$ subspace, 180$\mu$s for two-dimensional maps $\mathcal{H}_i \to \mathcal{H}_f$, and 100$\mu$s for state-to-state maps (Fig. 2).

So far the focus has been on optimization for a given physical system and laboratory setup, in the presence of the imperfections and perturbations that remain after concerted efforts to eliminate them. To learn more about the prospects for control in less benign environments, we can study the performance of robust control waveforms in the presence of much larger, deliberately introduced perturbations. As an example, consider static and dynamic variations in the bias field, $B(t) = B_0 + \delta B(t)$. In our case $\delta B(t)$ is dominated by the 60Hz power line cycle, and thus any change during control times $T \le 1$ ms will be approximately linear. This situation is typical of many cold-atom experiments, but more complex time variations can presumably be addressed with advanced decoupling schemes[23].

Figure 5 shows theoretically predicted fidelities for unitary maps in the presence of perturbations $\delta B(t) = \delta B_i + (\delta B_f - \delta B_i)\, t / T$, conveniently characterized by the initial and final values of the bias field variation. Non-robust waveforms were designed to maximize the fidelity only at the nominal bias field (Fig. 5a), resulting in poor performance for even small $\delta B_i, \delta B_f$ (Fig. 5b). Robust waveforms, by contrast, were designed to maximize the average fidelity for four different situations: static offsets $\delta B_i = \delta B_f$, and linear variations $\delta B_i = -\delta B_f$ (Fig. 5c). This improves the fidelity significantly for a wide range of static and dynamical perturbations (Fig. 5d), expanding, e. g., the 0.99 fidelity contour by roughly a factor of five compared to non-robust waveforms. The tradeoff is a control time $T = 800\,\mu s$, about 35% longer than non-robust waveforms.

To verify the performance of robust and non-robust waveforms in the laboratory, we perform randomized benchmarking of unitary maps at several points along the $\delta B_i = \delta B_f$ and $\delta B_i = -\delta B_f$ diagonals. As shown in Fig. 5e&f, the predicted and observed increases in robustness agree reasonably well. Note also that in the absence of a deliberately applied perturbation, $\delta B_i = \delta B_f = 0$, the robust control waveforms achieve the same peak fidelity as in Fig. 3. This is strong evidence that residual dynamic perturbations are insignificant in our setup.

Looking ahead, one immediate issue is how to increase fidelity in our setup, whether by improving the accuracy of our control fields, or by further reducing external perturbations. It will also be advantageous to shorten control times, either by increasing the strength of our control fields, or by adding, e. g, a second $\mu$w field to couple the states $|F^{(\pm)}, m = -F^{(\pm)}\rangle$. In the long term, there are a number of important questions still to explore. What are the practical limits on optimal control, and will this permit accurate and robust control of less ideal systems, e. g., atoms in optical traps? How large a Hilbert space can one realistically hope to control by the means used here? And how do the answers to these and other questions depend on the structure of the control Hamiltonian, notably its connectedness[22]? Can inhomogeneous control[6] be extended to qudits, perhaps allowing addressable unitary maps on qudits in a large array[24]? And finally, is it possible to optimize control in the presence of decoherence[25], and perhaps extend quantum control to include (non-unitary) completely positive maps[26]? Some of these questions can be explored in our current system, while others await the application of optimal control techniques to scalable architectures of interacting qubits and qudits.

## Methods

### Control Hamiltonian

The hyperfine Hamiltonian for a $^{133}$Cs atom in the presence of a magnetic field is of the form $H = A\mathbf{S}\cdot\mathbf{I} + g_S\mu_B\mathbf{S}\cdot\mathbf{B}(t) + g_I\mu_B\mathbf{I}\cdot\mathbf{B}(t)$, where $\mathbf{S}$ and $\mathbf{I}$ are the electron and nuclear spins, $g_S$ and $g_I$ the g-factors, and $\mathbf{B}(t) = B_0\mathbf{z} + B_{rf}^{(x)}(t)\mathbf{x} + B_{rf}^{(y)}(t)\mathbf{y} + \mathbf{B}_{\mu w}(t)$. In the rotating wave approximation we obtain a control Hamiltonian of the form

$$H_C(t) = H_0 + H_{rf}^{(+)}[\phi_x(t), \phi_y(t)] + H_{rf}^{(-)}[\phi_x(t), \phi_y(t)] + H_{\mu w}[\phi_{\mu w}(t)] .$$

Here $H_0$ is a drift term that includes the hyperfine interaction and Zeeman shift from the bias field, $H_{rf}^{(\pm)}$ generate $SU(2)$ rotations of the $F^{(\pm)}$ hyperfine spins depending on the phases of the rf fields along $x$ and $y$, and $H_{\mu w}$ generate $SU(2)$ rotations of the $|F^{(\pm)}, m = F^{(\pm)}\rangle$ pseudospin depending on the $\mu$w phases. This is sufficient to make the coupled electron-nuclear spin system controllable[4]. Besides the control phases, $H_C(t)$ depends on an additional set of parameters $\Lambda$. In our setup these parameters and their nominal values are: the Larmor frequency of the $F^{(+)}$ spin in the bias field $(\Omega_0 = g_{F^{(+)}}\mu_B B_0/\hbar = 2\pi\times 1\text{ MHz} \Rightarrow B_0 \approx 3\text{ G})$, the rf Larmor frequencies in the rotating frame $(\Omega_x = \Omega_y = 2\pi\times 25\text{ kHz})$, the $\mu$w Rabi frequency $(\Omega_{\mu w} = 2\pi\times 27.5\text{ kHz})$, and the rf and $\mu$w detunings from resonance $(\Delta_{rf} = \Delta_{\mu w} = 0)$. For the detailed form of $H_C(t)$ see Refs. [27,28].

### Numerical design of control waveforms

The map $U(T)$ implemented by $H_C(t)$ is found by solving the Schrödinger equation, a process made more efficient by the use of piece-wise constant control phases. Once $U(T)$ is known the fidelity $\mathcal{F}_S$ relative to the target map is easily computed. When searching for robust control waveforms, this process is repeated for each value of the inhomogeneous parameter(s) in $\Lambda$ and the resulting fidelities averaged. In principle one can use standard MatLab optimization tools to maximize the fidelity with respect to the control phases, but in practice we have found it advantageous to supplement these with a numerically efficient algorithm to calculate gradients. We use a variant of the GRAPE algorithm[29] modified for non-infinitesimal phase step duration[30]. With this we find numerical optimization of individual control waveforms to be straightforward on a desktop computer, and the design of large numbers of robust control waveforms to be feasible on a high-performance cluster. For details, including MatLab code, see Ref. [28].

### Experimental setup

Our experimental setup[16] is built around a vapor cell magneto-optic trap (MOT) and optical molasses, used to prepare an ensemble of a few million atoms in free fall at $\mu$K temperatures. At the start of each experimental cycle the atoms are initialized by optical pumping into the state $|F^{(+)}, m = F^{(+)}\rangle$. Bias and rf magnetic fields are applied with orthogonal coil pairs driven by arbitrary waveform generators. The $\mu$w magnetic field is generated by a fixed-frequency $\mu$w synthesizer mixed with an arbitrary waveform generator, and radiated by two separate antennae adjusted to optimize spatial homogeneity of the $\mu$w power across the ensemble. Applied and background magnetic fields are measured using the atoms themselves as in-situ sensors, and the latter cancelled by adding compensating currents to the bias and rf coils. As a result, our combined bias and background fields are accurate to 20ppm and stable to 10ppm (30$\mu$G). Populations in the magnetic sublevels $|F, m\rangle$ are measured via Stern-Gerlach analysis, implemented by letting the atoms fall in the presence of a magnetic field gradient and recording their arrival time at a probe beam located below the MOT.

### Randomized benchmarking

To determine the average fidelity of a given class of quantum maps, we first design control waveforms for a randomly chosen, representative sample. These maps are then combined in random sequences

$$|F, m = F^{(+)}\rangle \rightarrow |\psi_0\rangle \xrightarrow{U_1} |\psi_1\rangle \xrightarrow{U_2} \ldots \xrightarrow{U_l} |\psi_l\rangle \rightarrow |F, m = F^{(+)}\rangle .$$

Each sequence of length $l$ begins with a state preparation step (optical pumping into $|F, m = F^{(+)}\rangle$ and a state-to-state map to the randomly chosen input $|\psi_0\rangle$), and ends with a measurement step to determine the overall fidelity $\mathcal{F}(l)$ (mapping from the expected output state $|\psi_l\rangle$ to $|F, m = F^{(+)}\rangle$ and measuring its population by Stern-Gerlach analysis). For each $l$ we average the results from 10 different sequences to smooth out variations from accidental spin-echo effects. The resulting data is fit to a function

$$\mathcal{F}(l) = \frac{1}{d} + \frac{d-1}{d}\left(1 - \frac{d}{d-1}\varepsilon_0\right)\left(1 - \frac{d}{d-1}\varepsilon_B\right)^l ,$$

where $\varepsilon_0$ is the combined state preparation and measurement (SPAM) error, and $\varepsilon_B = 1 - \mathcal{F}_B$ is the average error per map estimated by the benchmarking procedure. Numerical simulation of the benchmarking procedure in the presence of known imperfections show close correlation between the standard ($\varepsilon_S = 1 - \mathcal{F}_S$) and benchmark errors, $0.5\varepsilon_B < \varepsilon_S < 1.35\varepsilon_B$ for 16-dimensional unitary maps, and $0.5\varepsilon_B < \varepsilon_S < 1.15\varepsilon_B$ for state-to-state maps. See Refs. [16,28] for details.

## Acknowledgements

This work was supported by the US National Science Foundation Grants PHY-1212308, PHY-1212445, PHY-1307520.